\documentclass{article}
\usepackage{amsmath}
\usepackage{amsfonts}

\usepackage{amsmath,amssymb}
\usepackage{graphicx}
\usepackage{color}

\newtheorem{fact}{Fact}

\newtheorem{theorem}{Theorem}
\newtheorem{lemma}{Lemma}
\newtheorem{corollary}{Corollary}

\newtheorem{definition}{Definition}

\newtheorem{remark}{Remark}

\makeatletter

\newcommand{\Rmnum}[1]{\expandafter\@slowromancap\romannumeral #1@}
\makeatother

\newcommand{\done}{\hfill $\Box$ }

\newcommand{\ls}[1]
    {\dimen0=\fontdimen6\the\font\lineskip=#1\dimen0
     \advance\lineskip.5\fontdimen5\the\font
     \advance\lineskip-\dimen0
     \lineskiplimit=0.9\lineskip
     \baselineskip=\lineskip
     \advance\baselineskip\dimen0
     \normallineskip\lineskip\normallineskiplimit\lineskiplimit
     \normalbaselineskip\baselineskip
     \ignorespaces}

\begin{document}

\bibliographystyle{abbrv}

\title{A Generalized Construction of OFDM $M$-QAM Sequences With Low Peak-to-Average Power Ratio}
\author{Zilong Wang$^{*1,2}$, Guang Gong$^2$, and Rongquan Feng$^1$\\
$^1$ LMAM, School of Mathematical Sciences, Peking University, \\
Beijing 100871,  P.R. China\\
$^2$ Department of Electrical and Computer Engineering, University of Waterloo \\
Waterloo, Ontario N2L 3G1, Canada \\
Email: wzlmath@gmail.com \ \  ggong@calliope.uwaterloo.ca \ \ fengrq@math.pku.edu.cn\\
}

 \maketitle

\footnotetext[0] {The work is supported by NSERC Discovery Grant.
$^*$Zilong Wang is currently a visiting Ph. D student at the
Department of ECE in University of Waterloo from September 2008 to
Setember 2009. The third author is supported by NSF of China (No.
10990011).}

\thispagestyle{plain} \setcounter{page}{1}

\begin{abstract}
A construction of $2^{2n}$-QAM sequences is given and an upper bound
of the peak-to-mean envelope power ratio (PMEPR) is determined. Some
former works can be viewed as special cases of this construction.

\end{abstract}

\ls{1.5}
\section{Introduction}

Multicarrier communications have recently attracted much attention
in wireless applications. The orthogonal frequency division
multiplexing (OFDM) has been employed in several wireless
communication standards.  Their popularity is mainly due to the
robustness to multipath fading channels and the efficient hardware
implementation employing fast Fourier transform (FFT) techniques.
However, multicarrier communications have the major drawback of the
high peak-to-average power ratio (PAPR) of transmitted signals.
Please refer to Litsyn \cite{Litsyn1} for a general source on PAPR
control.

A coding method for PAPR control in multicarrier communications is
to use Golay complementary sequences \cite{Golay1} \cite{Golay2} for
subcarriers such that the sequences provide low peak-to-mean
envelope power ratio (PMEPR) of at most 2 for transmitted signals,
where the PAPR of the signals is bounded by the PMEPR. An important
theoretical research on Golay complementary sequences has been set
by Davis and Jedwab \cite{DJ}, where they showed the sequences can
be constructed as a coset of the first order Reed-Muller codes by
using algebraic normal forms. The research on Golay sequences has
been flourished in the literature \cite{FJP}, \cite{Paterson}. The
reader is referred to Jedwab \cite{Jedwab} for a comprehensive
survey of Golay sequences.

The approaches above consider phase-shift keying (PSK) signal
constellations. However, there are many OFDM systems utilizing
quadrature amplitude modulation (QAM) constellations. Some
constructions of 16-QAM and 64-QAM complementary sequences were
presented sequentially by Chong {\it et. al} \cite{CVT}, Lee and
Golomb \cite{LG}, and Li \cite{Li}. In 2001, R\"{o}{\ss}ing and
Tarokh \cite{RT} gave an upper bound of PMEPR of the set of 16-QAM
sequences using 2 quaternary phase-shift keying (QPSK) Golay
sequences. In 2003, Tarokh and Sadjadpour \cite{TS} generalized the
results in \cite{RT} from 16-QAM to $2^{2n}$-QAM sequences by using
$n$ QPSK Golay sequences, and determined an upper bound of the PMEPR
of this set. Motivated by these works, we found that the former
construction can be generalized in a new way, such that the family
size is significantly enlarged while the upper bound of PMEPR
changes insignificantly.

The rest of the paper is organized as follows. In Section 2, the
mathematical model of the multicarrier communication, the basic
concept of Golay sequences, and the main results in \cite{RT} and
\cite{TS} are reviewed. In Section 3, we give a construction of
$2^{2n}$-QAM sequences set $\mathcal{A}$, and determine an upper
bound of PMEPR($\mathcal{A}$). The construction in \cite{RT} and
\cite{TS} can be viewed as a special case of our construction.
Section 4 is for discussions and conclusions of this construction.

\section{Preliminaries}
\subsection{Definitions}
The transmitted OFDM signal is the real part of the complex signal
$$S_{\mathbf{a}}(t)=\sum_{i=0}^{N-1}a_ie^{2j\pi f_i t},$$
where $f_i$ is the frequency of the $i$th carrier, $j=\sqrt{-1}$,
and $\mathbf{a}=(a_1, a_2, \cdots, a_{N-1})$ is a sequence with
period $N$. To ensure orthogonality of different carriers, the $i$th
carrier frequency $f_i$ is set to be $f_0+i\triangle f$, where $f_0$
is the smallest carrier frequency and $\triangle f$ is an integer
multiple of the OFDM symbol rate $1/T$, namely, $T\triangle f\in
\mathbb{Z}$.

\begin{definition}
The instantaneous envelope power of $S_{\mathbf{a}}(t)$ is defined
as $P_{\mathbf{a}}(t)=|S_{\mathbf{a}}(t)|^2$. Then
$$P_{\mathbf{a}}(t)=S_{\mathbf{a}}(t)\cdot
\overline{S_{\mathbf{a}}(t)}=\sum_{i=0}^{N-1}\sum_{k=0}^{N-1}a_i\overline{a_k}
e^{2j\pi (i-k)\triangle ft}.$$
\end{definition}
Thus, the mean power of $S_{\mathbf{a}}(t)$ during the symbol period
$T$ is
$$\frac{1}{T}\int_0^T
P_{\mathbf{a}}(t)dt=\sum_{i=0}^{N-1}|a_i|^2=\|\mathbf{a}\|^2.$$

\begin{definition}
The peak envelope power (PEP) of a codeword $\mathbf{a}$ is defined
as $\mbox{PEP}(\mathbf{a})=sup_{t\in [0, T]}P_{\mathbf{a}}(t)$.
\end{definition}

\begin{definition}
The peak-to-mean power ratio (PMEPR) of a code $\mathcal{C}$ is
defined as
$$\emph{PMEPR}(\mathcal{C})=\max_{\mathbf{a}\in\mathcal{C}}\emph{PEP}(\mathbf{a})/P_{av}(\mathcal{C}),$$
where $P_{av}(\mathcal{C})$ is the mean envelope power of an OFDM
signal averaged over all the OFDM signals in the codebook
$\mathcal{C}$, i.e.,
$$P_{av}(\mathcal{C})=\frac{1}{T}\sum_{\mathbf{a}\in\mathcal{C}}p(\mathbf{a})
\int_0^T
P_{\mathbf{a}}(t)dt=\sum_{\mathbf{a}\in\mathcal{C}}p(\mathbf{a})\|
\mathbf{a} \|^2.$$
\end{definition}

\subsection{Golay sequences}
An $H$-ary PSK ($H$-PSK) constellation can be realized as
$\{e^{\frac{2\pi js_i}{H}} | s_i\in \mathbb{Z}_H\}$. Thus any
$H$-PSK sequence $\mathbf{a}=(a_0, a_1, \cdots, a_{N-1})$ is
associated with the sequence $\mathbf{s}=(s_0, s_1, \cdots,
s_{N-1})$, where $a_i=e^{\frac{2\pi js_i}{H}}$. For a given $H$-PSK
code $\mathcal{C}$, since $P_{av}(\mathcal{C})=N$, the PMEPR of code
$\mathcal{C}$ can be determined as
$$\mbox{PMEPR}(\mathcal{C})=\max_{\mathbf{a}\in \mathcal{C}
}\mbox{PEP}(\mathbf{a})/N.$$ By the definition, one can get
$N\leqslant\mbox{PEP}(\mathbf{a})\leqslant N^2$. Thus
$1\leqslant\mbox{PMEPR}(\mathcal{C}) \leqslant N$.

An efficient coding method to reduce the PMEPR to 2 is the Golay
sequences which was first introduced by M. J. E. Golay \cite{Golay1}
in the context of infrared spectrometry. This approach relegates the
main difficulty of reducing the PMEPR from finding the flat
polynomials to constructing the sequences with good aperiodic
auto-correlation property, i.e., from a continuous problem to a
discrete one.
\begin{definition}
The aperiodic auto-correlation of the sequence $\mathbf{a}=(a_1, a_2
\cdots, a_{N-1})$ at shift $\tau$, where $1\leqslant\tau\leqslant
N-1$, is defined as
$$C_{\mathbf{a}}(\tau)=\sum_{i=0}^{N-1-\tau}a_i\overline{a_{i+\tau}}.$$
\end{definition}
Thus $P_{\mathbf{a}}(t)$ can be rewritten as the form
\begin{eqnarray*}
P_{\mathbf{a}}(t)&=&\sum_{i=0}^{N-1}\sum_{k=0}^{N-1}a_i\overline{a}_k
e^{2j\pi (i-k)\triangle ft}\\
&=&\sum_{i=0}^{N-1}|a_i|^2+2\sum_{\tau=1}^{N-1}\mbox{Re}(e^{-2j\pi\tau\Delta
ft}\sum_{i=0}^{N-1-\tau}a_i\overline{a_{i+\tau}})\\
&=&N+2\sum_{\tau=1}^{N-1}\mbox{Re}(e^{-2j\pi\tau\Delta
ft}C_{\mathbf{a}}(\tau)).
\end{eqnarray*}
If a pair of sequences $\mathbf{a}$ and $\mathbf{b}$ satisfy
$$C_{\mathbf{a}}(\tau)+C_{\mathbf{b}}(\tau)=0,\ \ \ \forall \tau\neq 0,$$
then $P_{\mathbf{a}}(t)+P_{\mathbf{b}}(t)=2N$. This implies that
both $\mbox{PEP}(\mathbf{a})$ and $\mbox{PEP}(\mathbf{b}) \leqslant
2N$. Therefore, the PMEPR of the code $\mathcal{C}$, which is a
collection of these sequences, is not larger than 2.
\begin{definition}
The pair $(\mathbf{a}, \mathbf{b})$ satisfying the above condition
is called a {\em Golay complementary pair}. Each member of a Golay
complementary pair is called a {\em Golay complementary sequence},
or simply {\em Golay sequence}.
\end{definition}

\subsection{$M$-QAM sequences constructed from QPSK Golay sequences}
The QPSK constellation can be realized as $\{j^{m} \mid m\in
\mathbb{Z}_4\}$. Therefore the QPSK sequence $\mathbf{a}=(a_0, a_1,
\cdots, a_{N-1})$ is corresponding to the sequence $\mathbf{s}=(s_0,
s_1, \cdots, s_{N-1})$, where $a_i=j^{s_i}$ with $s_i\in
\mathbb{Z}_4$, and a $2^{2n}$-QAM constellation can be realized as
$$2^{2n}\mbox{-QAM}=\sum_{i=0}^{n-1}2^{n-1-i}\frac{\sqrt{2}}{2}j^{s_i}e^{\frac{\pi j}{4}}=\frac{\sqrt{2}}{2}e^{\frac{\pi
j}{4}}\sum_{i=0}^{n-1}2^{n-1-i}j^{s_i}.$$  $2^4$-QAM constellation
can be viewed in both \cite{RT} and \cite{CVT} as a simple example
when $n=2$. In this way, any $2^{2n}$-QAM sequence $\mathbf{a}=(a_0,
a_1, \cdots, a_{N-1})^T$ with period $N$ is associated with a
sequence vector or a matrix $\mathbf{s}=(\mathbf{s_0}, \mathbf{s_1},
\cdots, \mathbf{s_{n-1}})$, where $\mathbf{s_i}=(s_{i,0}, s_{i,1},
\cdots, s_{i,N-1})^T\in \mathbb{Z}_4^N$ is a quaternary sequence
with period $N$. In particular, the $k$th element of the
$2^{2n}$-QAM sequence $\mathbf{a}$ is associated with $(s_{1,k},
s_{2,k}, \cdots, s_{{n-1},k})$, and can be presented as
$$a_k=\frac{\sqrt{2}}{2}e^{\frac{\pi
j}{4}}\sum_{i=0}^{n-1}2^{n-1-i}j^{s_{i,k}}.$$ Thus the signal
$S_{\mathbf{a}}(t)$ can be written as
$$S_{\mathbf{a}}(t)=\frac{\sqrt{2}}{2}\sum_{k=0}^{N-1}\sum_{i=0}^{n-1}2^{n-1-i}j^{s_{i,k}}e^{2\pi jf_k t+\frac{\pi
j}{4}}.$$

Let $\mathcal{C}$ be a collection of the $2^{2n}$-QAM sequences
$\mathbf{a}$ corresponding to $\mathbf{s}=(\mathbf{s_0},
\mathbf{s_1}, \cdots, \mathbf{s_{n-1}})$, where $\mathbf{s_i}$ is a
Golay sequence for any $0\le i\le n-1$. An upper bound of
PMEPR($\mathcal{C}$) is determined in \cite{RT} for $16$-QAM and in
\cite{TS} for the general case, which is shown as follows.
\begin{fact}
$$\emph{PMEPR} (\mathcal{C})\leqslant \frac{6(2^n-1)^2}{2^{2n}-1}.$$
\end{fact}

From Fact 1, it's straightforward to get that $\mbox{PMEPR}
(\mathcal{C})\leqslant 3.6$ for $16$-QAM, and $\mbox{PMEPR}
(\mathcal{C})< 6$ for general $n$.

\section{A generalized construction with low PMEPR}
For two given numbers $x, y$ with $x>1$ and $1\leqslant y<2$, let
$\mathcal{S}_i$ ($0\leqslant i\leqslant n-1$) be a subset of the
QPSK sequences with period $n$, and satisfy the following
conditions:
\begin{enumerate}
\item[(a)] ${\rm PEP}(\mathbf{s_i})\leqslant xy^{2i}N$ for every
$\mathbf{s_i}\in \mathcal{S}_i$.
\item[(b)] If $\mathbf{s_i}\in \mathcal{S}_i$, then $j^{m}\mathbf{s_i}\in
\mathcal{S}_i$ for $m\in \mathbb{Z}_4$, where
$j^{m}\mathbf{s_i}=(j^ms_{i,0}, j^ms_{i,1}, \cdots, j^ms_{i,n-1})$.
\end{enumerate}
\begin{remark}

\begin{enumerate}
\em{
\item[1)] It is not required that $\mathcal{S}_i$ contains all the
sequences satisfying ${\rm PEP}(\mathbf{s_i})\leqslant xy^{2i}N$.
\item[2)] ${\rm PEP}(\mathbf{s_i})={\rm PEP}(j^m\mathbf{s_i})$}, so it is
reasonable to require $\mathcal{S}_i$ satisfy the condition (b).
\end{enumerate}
\end{remark}

\begin{theorem}
Let $\mathcal{A}$ be a collection of the $2^{2n}$-QAM sequences
$\mathbf{a}$ such that $\mathbf{a}=(a_0, a_1, \cdots,
a_{N-1})^T=(\mathbf{s_0}, \mathbf{s_1}, \cdots, \mathbf{s_{n-1}})$
and $\mathbf{s_i}\in \mathcal{S}_i$. Then
$$\emph{PMEPR} (\mathcal{A})\leqslant \frac{3}{4}\cdot \frac{2^{2n}}{2^{2n}-1}\cdot
\left(\frac{1-(\frac{y}{2})^{n}}{1-\frac{y}{2}} \right)^2 \cdot x.$$
\end{theorem}

For verifying Theorem 1, we first estimate PEP$(\mathbf{a})$ for
every $\mathbf{a} \in \mathcal{A}$ in Lemma 1, then determine
$\emph{P}_{av}(\mathcal{A})$ in Lemma 2.
\begin{lemma}
Let $\mathbf{a}$ be a $2^{2n}$-QAM sequence such that
$\mathbf{a}=(a_0, a_1, \cdots, a_{N-1})^T=(\mathbf{s_0},
\mathbf{s_1}, \cdots, \mathbf{s_{n-1}})$ and $\mathbf{s_i}\in
\mathcal{S}_i$. Then
$$\emph{PEP}(\mathbf{a})\leqslant 2^{2n-3}\left(\frac{1-(\frac{y}{2})^{n}}{1-\frac{y}{2}} \right)^2\cdot x\cdot N.$$
\end{lemma}
{\em Proof}: The signal $S_{\mathbf{a}}(t)$ can be written in the
form
\begin{eqnarray*}
S_{\mathbf{a}}(t)&=&\frac{\sqrt{2}}{2}\sum_{k=0}^{N-1}\sum_{i=0}^{n-1}2^{n-1-i}j^{s_{i,k}}e^{2\pi
jf_k t+\frac{\pi j}{4}}\\
&=&\frac{\sqrt{2}}{2}e^{\frac{\pi
j}{4}}\sum_{i=0}^{n-1}2^{n-1-i}\sum_{k=0}^{N-1}j^{s_{i,k}}e^{2\pi
jf_k t}\\
&=&\frac{\sqrt{2}}{2}e^{\frac{\pi
j}{4}}\sum_{i=0}^{n-1}2^{n-1-i}S_{\mathbf{s_i}}(t).
\end{eqnarray*}
Thus the instantaneous envelope power of $\mathbf{a}$ is given by
$$P_{\mathbf{a}}(t)=|S_{\mathbf{a}}(t)|^2=\frac{1}{2}\left|\sum_{i=0}^{n-1}2^{n-1-i}S_{\mathbf{s_i}}(t)\right|^2.$$
By the triangle inequality, one can get
$$P_{\mathbf{a}}(t)\leqslant \frac{1}{2}\left(\sum_{i=0}^{n-1}2^{n-1-i}|S_{\mathbf{s_i}}(t)|\right)^2.$$
From $\mathbf{s_i} \in \mathcal{S}_i$ and ${\rm
PEP}(\mathbf{s_i})\leqslant xy^{2i}N$, we have
$|S_{\mathbf{s_i}}(t)|\leqslant (xy^{2i}N)^{\frac{1}{2}}$. Thus
\begin{eqnarray*}
P_{\mathbf{a}}(t)&\leqslant&
\frac{1}{2}\left(\sum_{i=0}^{n-1}2^{n-1-i}\left(xy^{2i}N\right)^{\frac{1}{2}}\right)^2\\
&=&\frac{1}{2}x N \left(\sum_{i=0}^{n-1}2^{n-1-i}y^i\right)^2\\
&=&\frac{1}{2}x N \left(2^{n-1}\sum_{i=0}^{n-1}\left(\frac{y}{2}\right)^i\right)^2\\
&=&2^{2n-3}\left(\frac{1-(\frac{y}{2})^{n}}{1-\frac{y}{2}}
\right)^2\cdot x\cdot N
\end{eqnarray*} \done

\begin{lemma}
Let $\mathbf{a}$ be a $2^{2n}$-QAM sequence such that
$\mathbf{a}=(a_0, a_1, \cdots, a_{N-1})^T=(\mathbf{s_0},
\mathbf{s_1}, \cdots, \mathbf{s_{n-1}})$ and $\mathbf{s_i}\in
\mathcal{S}_i$. Then
$$\emph{P}_{av}(\mathcal{A})=\frac{1}{2}(2^n-1)\cdot N.$$
\end{lemma}
{\em Proof}: Regard $\mathbf{a}$ as a discrete random variable such
that every $\mathbf{s_i}$ is chosen from $\mathcal{S}_i$ with the
same probability, as well as the time $t$ is a continuous random
variable uniformly distributed in the interval $[0, T]$. Then
$\emph{P}_{av}$ can be regarded as the expectation of the random
function $P_{\mathbf{a}}(t).$ In the following, we also treat the
sequence $\mathbf{s_i}$, and $s_{i,j}$, the $j$th element of
$\mathbf{s_i}$, as random variables. Therefore
\begin{eqnarray*}
P_{av}(\mathcal{A})&=&E(P_{\mathbf{s}}(t))\\
&=&E\left(\frac{1}{2}\left|\sum_{i=0}^{n-1}2^{n-1-i}S_{\mathbf{s_i}}(t)\right|^2\right)\\
&=&\frac{1}{2}E\left(\sum_{i=0}^{n-1}2^{n-1-i}S_{\mathbf{s_i}}(t)\cdot
\sum_{k=0}^{n-1}2^{n-1-k}\overline{S_{\mathbf{s_k}}(t)}\right)\\
&=&\frac{1}{2}\sum_{i=0}^{n-1}\sum_{k=0}^{n-1}2^{2n-2-i-k}E\left(S_{\mathbf{s_i}}(t)\overline{S_{\mathbf{s_k}}(t)}\right)\\
&=&\sum_{i=0}^{n-1}2^{2n-3-2i}E|S_{\mathbf{s_i}}(t)|^2+\frac{1}{2}\sum_{i=0}^{n-1}\sum_{k\neq
i}2^{2n-2-i-k}
E\left(S_{\mathbf{s_i}}(t)\overline{S_{\mathbf{s_k}}(t)}\right).\\
\end{eqnarray*}
Since $\mathbf{s_i}$ is a random variable with respect to QPSK
sequences in $\mathcal{S}_i$, one can get
$E|S_{\mathbf{s_i}}(t)|^2=N$ immediately. For $k\neq i$,
\begin{eqnarray*}
E\left(S_{\mathbf{s_i}}(t)\overline{S_{\mathbf{s_k}}(t)}\right)&=&E\left(\sum_{p=0}^{N-1}s_{i,p}e^{2j\pi
(f_0+p\Delta f)t}\sum_{q=0}^{N-1}\overline{s_{k,q}}e^{-2j\pi
(f_0+q\Delta
f)t}\right)\\
&=&\sum_{p=0}^{N-1}\sum_{q=0}^{N-1}E\left(s_{i,p}\overline{s_{k,q}}e^{2j\pi
(p-q)\Delta f t}\right).
\end{eqnarray*}
For given $i, k, p, q$ with $i\neq k$, $s_{i,p}$ and $s_{k,q}$ are
random variables with respect to the $p$th and $q$th elements of
$\mathbf{s_i}$ and $\mathbf{s_k}$ respectively. So $s_{i,p}$ and
$s_{k,q}$ are independent. Thus,
$$E\left(s_{i,p}\overline{s_{k,q}}e^{2j\pi
(p-q)\Delta f
t}\right)=E(s_{i,p})\overline{E(s_{k,q})}E\left(e^{2j\pi (p-q)\Delta
f t}\right).$$ By the definition, if a sequence $\mathbf{s_i}\in
\mathcal{S}_i$, then $j^{m}\mathbf{s_i}\in \mathcal{S}_i$. Therefore
$s_{i,p}=j^m$ with the equal probability $1/4$ for any $m\in
\mathbb{Z}_4$, which implies $E(a_{i_p})=0$. Due to the above, we
obtain
$$P_{av}(\mathcal{A})=N\sum_{i=0}^{n-1}2^{2n-3-2i}=\frac{N}{6}(2^{2n}-1).$$
This completes the proof. \done

{\em Proof of Theorem 1:} By the results of Lemmas 1 and 2, the
assertion of Theorem 1 follows immediately from the definition of
PMEPR. \done

\begin{corollary}
$$\emph{PMEPR} (\mathcal{A})< \frac{3}{4}\cdot \frac{x}{(1-\frac{y}{2})^2}.$$
\end{corollary}

{\em Proof:} For $1\leqslant y< 2$, it is obvious that
$$\lim_{n\rightarrow +\infty}\frac{2^{2n}}{2^{2n}-1}\cdot
\left(1-\left(\frac{y}{2}\right)^n\right)^2=1.$$ Thus, to prove the
result, one needs only to verify $\frac{2^{2n}}{2^{2n}-1}\cdot
(1-(\frac{y}{2})^n)^2$ is an increasing function with respect to $n$
when $n\geqslant 1$. From
$$\frac{2^{2n}}{2^{2n}-1}\cdot
\left(1-\left(\frac{y}{2}\right)^n\right)^2=\left(1-\left(\frac{y}{2}\right)^n\right)\cdot
\left(1-\frac{1}{2^n+1}\right)\cdot
\left(1-\frac{y^n-1}{2^n-1}\right),$$ the claim holds since all
$1-\left(\frac{y}{2}\right)^n$, $1-\frac{1}{2^n+1}$, and
$1-\frac{y^n-1}{2^n-1}$ are positive increasing functions with
respect to $n$ when $n\geqslant 1$ and $1\leqslant y< 2$. This
completes the proof.  \done

\begin{corollary}
Let $y=1+\epsilon\ (\epsilon\geqslant 0)$, then
$$\emph{PMEPR} (\mathcal{A})<3x(1+2\epsilon)+o(\epsilon)\ \ \  \emph{and}\ \ \ \ \emph{PMEPR} (\mathcal{A})< 3xy^2+o(\epsilon).$$
\end{corollary}

{\em Proof:} Since $y=1+\epsilon$, we have
$$\frac{3}{4}\cdot \frac{x}{(1-\frac{y}{2})^2}=\frac{3x}{(1-\epsilon)^2}=3x(1+2\epsilon)+o(\epsilon)$$
and $$3xy^2=3x(1+\epsilon)^2=3x(1+2\epsilon)+o(\epsilon).$$ \done

\section{Conclusion}

Note that Fact 1 in Section 2.3, the main result in \cite{TS} and in
\cite{RT}, can be viewed as a special case of Theorem 1 by setting
$x=2$ and $y=1$.

In the following, we discuss the case $y>1$.

First, we consider the QPSK sequences subset $\mathcal{S}$ with
${\rm PEP}(\mathbf{s})\leqslant \delta$ for all $\mathbf{s}\in
\mathcal{S}$. Obviously, there is a trade off between the size
$\#(\mathcal{S})$ and the upper bound $\delta$ of the set
$\mathcal{S}$. Since $\delta=xy^{2i}$ which may be larger than 2,
one can construct $\mathcal{S}_i$ as a larger set than the Golay
sequences set. There has been some research on how to enlarge the
family size at the cost of increasing the PEP bound. The reader is
referred to \cite{PT} and \cite{Schmidt} for the construction of
near-complementary sequences with ${\rm PMEPR}<\delta$, and
\cite{SH}, \cite{Litsyn2}, and \cite{STH} for the construction of
$\mathcal{S}$ with family size $2^{cn}$ and PEP upper bound $c\log
n$.

Since $xy^{2i}$ is an exponential function with respect to $i$,
there exists $i_0$ such that $xy^{2i}\geqslant N$ when $i\geqslant
i_0$. This implies that the sequences in the set $S_i$ can be
arbitrary.

If $x=2$ and $y=1+\epsilon$ with a small number $\epsilon$, compared
with the set $\mathcal{C}$ presented in Fact 1, ${\rm
PMEPR}(\mathcal{A})$ changes insignificantly by Corollary 2, while
the size of the set $\mathcal{A}$ is significantly enlarged from the
above results.

From Corollary 2, ${\rm PMEPR}(\mathcal{A})$ is bounded by $3\cdot
{\rm PEP}(\mathcal{S}_1)$ if $\epsilon$ is small enough. An
interesting idea is that if there exist $x$ and $y$ with $xy^2<2$
and $\mathcal{S}_0$ is not an empty set, then one can obtain the
bound ${\rm PMEPR}(\mathcal{A})<6$. Here the size
$\#(\mathcal{S}_0)$ and $\#(\mathcal{S}_1)$ may be small, but
$\#(S_i)$ would be very large for large enough $i$ due to the
comments above, which ensures that $\mathcal{A}$ is a set with great
size.

\section*{Acknowledgment}
The authors would like to thank the anonymous referees for
suggestions. The first author is grateful to Qi Chai and Hong Wen
for fruitful discussions.

\end{document}